\documentclass[twocolumn]{aastex631}
\usepackage{natbib}

\usepackage{amsmath}
\DeclareUnicodeCharacter{FF1A}{:}
\DeclareUnicodeCharacter{FF1A}{-}

\begin{document}

% \title{Detecting high-energy neutrinos from Individual and Stacked GRB Sources Using Enhanced Neutrino Detectors}
\title{Prospect of Gamma-Ray Burst Neutrino Detection with Enhanced Neutrino Detectors}

\author{Wenkang Lian}
% \email{lianwenkang@mail.bnu.edu.cn}
\affiliation{School of Physics and Astronomy, Beijing Normal University, Beijing 100875, People's Republic of China}

\author[0000-0002-9165-8312]{Shunke Ai}
\affiliation{Niels Bohr International Academy and DARK, Niels Bohr Institute, University of Copenhagen, Blegdamsvej 17, 2100, Copenhagen, Denmark}

\author[0000-0003-2516-6288]{He Gao}
\affiliation{School of Physics and Astronomy, Beijing Normal University, Beijing 100875, People's Republic of China}
\affiliation{Institute for Frontier in Astronomy and Astrophysics, Beijing Normal University, Beijing 102206, People's Republic of China}

\correspondingauthor{Shunke Ai}
\email{shunke.ai@nbi.ku.dk}

\correspondingauthor{He Gao}
\email{gaohe@bnu.edu.cn}

%\collaboration{20}{(AAS Journals Data Editors)}
%
%\author{F.X Timmes}
%\affiliation{Arizona State University}
%\affiliation{AAS Journals Associate Editor-in-Chief}

%\author{Amy Hendrickson}
%\altaffiliation{AASTeX v6+ programmer}
%\affiliation{TeXnology Inc.}

%\author{Julie Steffen}
%\affiliation{AAS Director of Publishing}
%\affiliation{American Astronomical Society \\
%1667 K Street NW, Suite 800 \\
%Washington, DC 20006, USA}

%% Note that the \and command from previous versions of AASTeX is now
%% depreciated in this version as it is no longer necessary. AASTeX 
%% automatically takes care of all commas and "and"s between authors names.

%% AASTeX 6.31 has the new \collaboration and \nocollaboration commands to
%% provide the collaboration status of a group of authors. These commands 
%% can be used either before or after the list of corresponding authors. The
%% argument for \collaboration is the collaboration identifier. Authors are
%% encouraged to surround collaboration identifiers with ()s. The 
%% \nocollaboration command takes no argument and exists to indicate that
%% the nearby authors are not part of surrounding collaborations.

%% Mark off the abstract in the ``abstract'' environment. 
\begin{abstract}

%Observing neutrinos is an effective way to confirm the distance of the gamma-ray burst (\GRB) prompt emission region from the central engine, which helps determine which model—whether it is the dissipation photospheric, internal shock, or \ICMART\ model—produces the prompt emission. 
Gamma-ray bursts (GRBs) have long been proposed as a potential source of high-energy neutrinos. Although no confirmed association between GRBs and neutrinos has been established, meaningful constraints have been placed on GRB prompt emission models. The nondetection of neutrinos, reported by the IceCube Collaboration, from both single and stacked GRB events suggests that the radiation zone is likely located at a considerable distance from the central engine, where the photon number density is relatively low. Here, we estimate future GRB neutrino detection probabilities using detectors with a higher simulated sensitivity than IceCube and explore the constraints on models if GRB neutrinos remain undetected despite improved sensitivity. Our findings reveal that if the effective area of a future neutrino detector can be enhanced by a factor of 10 compared to IceCube IC86-II, there is a high likelihood of detecting neutrinos from a GRB 221009A-like event, even in the context of the ICMART model, which exhibits the lowest efficiency in neutrino production. With such an advanced detector (enhanced by a factor of 10) and 5–10 yr of data accumulation, neutrinos from stacked GRBs should be identifiable, or several popular models for GRB prompt emission (e.g., the dissipative photosphere model and internal shock model) could be effectively ruled out.

\end{abstract}

%% Keywords should appear after the \end{abstract} command. 
%% The AAS Journals now uses Unified Astronomy Thesaurus concepts:
%% https://astrothesaurus.org
%% You will be asked to selected these concepts during the submission process
%% but this old "keyword" functionality is maintained in case authors want
%% to include these concepts in their preprints.
%\keywords{Classical Novae (251) --- Ultraviolet astronomy(1736) --- History of %astronomy(1868) --- Interdisciplinary astronomy(804)}

%% From the front matter, we move on to the body of the paper.
%% Sections are demarcated by \section and \subsection, respectively.
%% Observe the use of the LaTeX \label
%% command after the \subsection to give a symbolic KEY to the
%% subsection for cross-referencing in a \ref command.
%% You can use LaTeX's \ref and \label commands to keep track of
%% cross-references to sections, equations, tables, and figures.
%% That way, if you change the order of any elements, LaTeX will
%% automatically renumber them.
%%
%% We recommend that authors also use the natbib \citep
%% and \citet commands to identify citations.  The citations are
%% tied to the reference list via symbolic KEYs. The KEY corresponds
%% to the KEY in the \bibitem in the reference list below. 

\section{Introduction} 
\label{sec:intro}
In the radiation zone of gamma-ray bursts (GRBs), a large number of gamma-ray photons are generated, either through thermal or nonthermal processes. Simultaneously, protons within the relativistic GRB jet are expected to be accelerated and acquire a relativistic random motion. Consequently, interactions between protons and gamma-ray photons, known as $p\gamma$ interactions, are unavoidable, leading to the production of high-energy neutrinos 
\citep[e.g.][]{zhang2018gamma}. 
This, as well as the interactions between baryons, make GRB events a compelling candidate as a potential source of high-energy neutrinos \citep[e.g.][]{Waxman_1997,Dermer_2003,Razzaque_2003,GUETTA2004429,Murase_2006_2,H_mmer_2012,Murase2013,Bustamante_2015,tamborra2015,refId0}.
%\textcolor{red}  { See studies like \cite{Waxman_1997,Rudolph_2023} and \cite{ou2024neutrinoconstraintsdetectionprospects} for a detailed discussion.}

The predicted flux of high-energy neutrinos generated by GRB events through $p\gamma$ interactions depends significantly on the model used to describe the GRB prompt emission \citep[e.g.][]{Zhang_2010,Pitik_2021,DeLia2024}. In the literature, three prominent models are often discussed: the dissipative photosphere model \citep{Rees_2005}, the internal shock model \citep{Rees_1994}, and the ICMART model \citep{Zhang_2010}. In these models, different characteristic radii for the radiation regions are proposed. These radii determine the gamma-ray photon number density, which in turn influences the rate of $p\gamma$ interactions and, consequently, affects the resulting neutrino flux. The dissipative photosphere model requires a matter-dominated GRB jet, in which the gamma-ray photons are released at the photosphere radius as $R_{\rm ph}\sim10^{11-12}~{\rm cm}$ \citep{Mészáros_2000,Pe’er_2007}. In the internal shock model, GRB photons are generated due to the collision of different layers inside the jet, which happens at a radius as $R_{\rm IS}\sim 10^{12-13}~{\rm cm}$ \citep{Rees_1994,Daigne_1998}. In the ICMART model, the GRB jet is assumed to be dominated by Poynting flux. The gamma-ray photons are generated at a much larger radius as $R_{\rm ICMART}\sim10^{15}~{\rm cm}$ where significant magnetic dissipation occurs \citep{Zhang_2010,McKinney_2011,Lazarian_2019}. 
%However, despite decades of research since the discovery of GRBs, the real mechanism behind their prompt emission remains unclear. Therefore, 
Joint observations of GRBs and neutrinos would offer an independent approach to test and distinguish between these competing models.

Theoretically, a considerable number of neutrinos could be produced by each GRB \citep{kimura2022}. However, the immense distance from the source, coupled with the current technological constraints of neutrino detectors, significantly limits our capacity to detect these neutrinos. IceCube is currently the most sensitive detector for astrophysical neutrinos, located beneath the ice layer at the geographic South Pole \citep{ice_2017}. 

However, even for GRB 221009A, the brightest-of-all-time burst, no associated neutrinos were detected by IceCube \citep{aiello2024searchneutrinoemissiongrb} . The IceCube Collaboration reported an upper limit on the neutrino flux associated with this event \citep{2022GCN.32665....1I}, leading to constraints on GRB models \citep{Murase_2022, Ai_2023,Rudolph_2023,2024arXiv240816748V,ou2024}. It is likely that the radiation radius of GRB 221009A is relatively far from the central engine, which is consistent with the ICMART model, featuring a jet dominated by magnetic fields \citep{Yang_2023}. The internal shock model remains viable if the jet has a relatively high bulk Lorentz factor \citep{Murase_2022, Ai_2023}, whereas the dissipative photosphere model is disfavored \citep{Ai_2023}.

Considering the difficulty of detecting neutrinos from a single source, stacked neutrino detection across multiple sources has become a more popular approach. As early as 2011, \cite{2011PhRvL.106n1101A} searched for neutrino emission within a 1 day time window before and after GRBs from 2008 to 2009. Later, in 2015, \cite{Aartsen_2015} analyzed the prompt emission phases of 506 GRBs from 2008 to 2012, placing constraints on the parameter space of the three models and estimating that GRBs could account for about 1\% of the total astrophysical neutrino flux. Then, in 2017, \cite{Aartsen_2017} extended the data range by 3 yr, including 1172 GRBs from 2008 to 2015. In 2022, \cite{2022ApJ...939..116A} further extended the search window to 14 days before or after GRBs to account for the contribution of afterglows to neutrino production. Also in 2022, \cite{Lucarelli_2023} used 10 yr of IceCube data to search for neutrinos during both the GRB prompt and afterglow phases. To date, this stacked detection approach has not yet succeeded in detecting GRB neutrinos.
%\textbf{and they ultimately constrained $\epsilon_p / \epsilon_e$ to below 10.}

Several near-future neutrino detectors have been proposed or are already under construction, such as IceCube Gen2 \citep{Aartsen_2021}, Baikal-GVD in Lake Baikal \citep{AVRORIN201130}, KM3NeT located in the Mediterranean Sea \citep{Adrián-Martínez_2016}  and those in the Pacific Ocean, like P-One \citep{Agostini_2020} and TRIDENT \citep{ye2024multicubickilometreneutrinotelescopewestern}. At this stage, it is of great interest to assess the prospects for detecting GRB neutrinos as detector sensitivity improves in the future. What is the detection probability for individual bright sources similar to GRB 221009A? After 5-10 yr of cumulative observations, what is the likelihood of detecting GRB neutrinos through stacked neutrino detection methods? If GRB neutrino detection remains elusive, how constraining would this be for existing models? This work will delve into these questions. To ensure the generality of our research conclusions, we do not focus on the sensitivity of any specific proposed detector but instead calculate by increasing the effective area of IceCube IC86-II by a certain factor.

In this work, the CGS unit system is adopted, and the convention $Q_x = Q/{10^x}$ is used, where $Q$ represents any physical quantity. This convention applies only when $x$ is a number.

%\section{HIGH-ENERGY NEUTRINO \FLUENCE OF \GRBS} \label{sec:two}
\section{Theoretical model} \label{sec:two}

In the GRB environment, gamma-ray photons are produced concurrently with the acceleration of electrons and baryons \citep{Waxman_1997}. These non-thermal protons interact with gamma-ray photons, as well as other protons and neutrons, producing high-energy neutrinos. Typically, the number density of photons is much higher than that of the baryons, making the $p\gamma$ interaction the dominant hadronic process.
% \citep[e.g.,][]{zhang2018gamma}. 
Here we focus on the $p\gamma$ process to generate high-energy neutrinos. In this section, we will introduce the theoretical models for GRB prompt emission and the $p\gamma$ neutrino production respectively.

\subsection{GRB prompt emission} \label{sec:two two}

The gamma-ray photons produced during the prompt emission of GRBs provide the source for $p\gamma$ interactions. The photon number density, as well as that of other particles, depends on the radius of the emission region, which increases as it gets closer to the central engine. The typical emission radius varies with different GRB models, thus makes the predicted neutrino flux model dependent. In this work, we include three popular GRB models: 

%In \GRB\ environment, the particle number density increases as you get closer to the central engine. A higher density of gamma-ray photons will further enhance the $p\gamma$ interaction, thereby producing more neutrinos. The prompt emission site is highly model dependent. Although \GRBentry is considered to be Poynting flux dominated, here we still discuss the three models of Internal shock(IS), Dissipative photosphere (ph) and \ICMART\ in order to make a comparison of the different models. 
\begin{itemize}

    \item \textit{Dissipative photosphere model.} The gamma-ray photons are generated and trapped in the opaque jet until it reaches the photosphere. The radius of the Thomson scattering photosphere can be estimated as $R_{\text{ph}} \simeq 3.7 \times 10^{11} {\rm cm} ~L_{w,52} \Gamma^{-3}_{2.5} $ \citep{Meszaros_2001}, where $\Gamma$ is the bulk Lorentz factor of the jet and $L_w = L_{\rm GRB}/\epsilon_e$. Here, we assume that a fraction $\epsilon_e$ of the dissipated energy in the jet goes into electrons, which then convert into gamma-ray photons.
    %\textcolor{red}{We assume that a fraction $\epsilon_e$ of the dissipated energy is transferred to electrons and fully radiated as gamma rays, a fraction $\epsilon_p$ is transferred to protons, and a fraction $\epsilon_B$ goes into the random magnetic field.}
    Shocks, neutron-proton collisions and magnetic reconnection are supposed to happen below the photosphere, so that protons are also expected to be accelerated \citep{Rees_2005,2007RSPTA.365.1171P,2008A&A...480..305G,Zhang_2010, Rudolph_2023}.

    \item \textit{Internal shock (IS) model.} The collision of different layers of a GRB jet would excite internal shocks, which can accelerate particles and emit gamma-ray photons. This could occur beyond the photosphere of gamma rays. The typical radius can be estimated as
    %The typical radius for the gamma-ray radiation could be estimated as 
    $R_{\rm IS}=2\Gamma^2 c \delta t_{\rm min}/\left(1+z\right)$ \citep{Rees_1994,Daigne_1998,zhang2018gamma}, where $\delta t_{\rm min}$ stands for the minimum variation timescale for the GRB light curve. Using $\delta t_{\rm min} \sim 0.1$ s and $\Gamma$ in order of 10-100, the photon emission and proton acceleration occur at a typical internal shock radius as about $10^{12}-10^{13}~ {\rm cm}$.

    \item \textit{ICMART model.} When the GRB jet is Poynting-flux dominated, the global magnetic field may remain undissipated beyond $R_{\rm IS} $ and $ R_{\rm ph}$. Internal collisions help to destroy the ordered magnetic fields, and a strong runaway magnetic dissipation process occurs at a large radius $ R_{\rm ICMART} \sim \Gamma^2 c \delta t_{\text{slow}} \sim 10^{15} \ \text{cm}$ \citep{Zhang_2010}, where $\delta t_{\text{slow}} \sim 1$ s is the slow variability component in the GRB light curves \citep{Gao_2012}. Particles, either directly accelerated in the reconnection zones or stochastically accelerated within the turbulent regions, emit synchrotron radiation photons, which power the observed prompt emission of GRBs \citep{Zhang_2010,Zhangbo_2014,Shao_2022}. 
    %This model have a much lower neutrino flux than IS and photosphere models.
    %the internal collisions and reconnection of the magnetic field at a larger radius may accelerate particles and produce gamma-ray photons \citep{zhang2013,zhangbo,shaoxy}. 
    %this model invokes a highly magnetized  outflow, which remains un-dissipated up to a large radius $R_{\rm ICMART} > R_{\rm IS} $ and $ R_{\rm ICMART} > R_{\rm ph}$ \citep{Zhang_2010}. 
    %Thus, this model have a much lower neutrino flux than IS or ph. Emission from the photosphere and internal shocks is greatly suppressed. Internal shocks help to destroy the ordered magnetic fields, and a strong run-away magnetic dissipation process occurs at a large radius $R_{\text{ICMART}} \sim \Gamma^2 \delta t_{\text{slow}} \sim 10^{15} \ \text{cm}$, where $\delta t_{\text{slow}} \sim 1s$ 1s is the slow variability component in the GRB  lightcurves\citep{Gao_2012}.
\end{itemize}

\subsection{high-energy neutrinos from GRBs} \label{sec:two one}
%the site of shocks or reconnection also could accelerate protons. 
%These non-thermal protons would interact with photos fields,magnetic fields,and other baryons,such as protons and neutrons, and generate neutrinos and photos. In these interactions, $p\gamma$and$pp/pn$ interactions could generate neutrinos.However,in \GRB\ environment,the number density of photos is usually much higher than that of protons or neutrons ,thus the $p\gamma$ mechanism is usually the dominant \hadronic\ interaction process.So,we will mainly discuss $p\gamma$ interactions\citep{zhang2018gamma}.

When high-energy protons interact with photons of proper energy, they would be in $\Delta$ resonance and produce $\Delta^+$.
Then, the $\Delta^+$ decays into mesons, which further decay into leptons and neutrinos \citep[e.g.][]{M_cke_1999}. The process can be described as
\begin{equation}
\resizebox{.9\hsize}{!}{$
p\gamma \to \Delta^+ \to 
\left\{
\begin{array}{ll}
n\pi^+ \to n\mu^+ \nu_\mu \to ne^+ \nu_e \bar{\nu}_\mu \nu_\mu, & \text{fraction } \frac{1}{3}, \\
p\pi^0 \to p\gamma, & \text{fraction } \frac{2}{3}.
\end{array}
\right.
$}
\end{equation}

Besides $ \Delta$-resonance, direct pion production and multiple-pion production channels can produce $\pi^+$, whose cross-section reaction is only a factor of a few smaller; thus, the contributions from them cannot be ignored.
% The cross section above the $ \Delta$-resonance regime is only a factor of a few smaller, thus the contributions from above the $ \Delta$-resonance cannot be ignored.
When direct pion production and multiple-pion production channels are considered, the portions of producing $\pi^+$and $\pi^0$ become 1/2 and 1/2, respectively \citep{zhang_2013,zhang2018gamma}.

%We start from the perspective of \fluence\ to normalize the neutrino spectrum with the total proton fluence
We follow \cite{Waxman_1997} and \cite{kimura2022} to calculate the predicted neutrino fluence, which can be written as
%Based on \citep{Waxman_1997} and \citep{kimura2022}, we can conclude the predicted neutrino fluence:
\begin{equation}
\begin{aligned}
\phi_{\nu}(E_{\nu}) 
%&= \frac{E_{\nu}^2 n_{\nu}(E_{\nu}) \left(1+z\right)}{4\pi D_L^2} \\
&=\frac{1}{8} f_{p\gamma} f_{\rm cooling} \frac{\left(\epsilon_p/\epsilon_e\right) S_\gamma}{\ln\left(E_{p,\text{max}}/E_{p,\text{min}}\right)}
\end{aligned}
\label{eq:neutrino_flux}
\end{equation}
where 
%$D_L = D_c \left(1+z\right)$ is the luminosity distance, 
$S_\gamma$ is the gamma ray fluence that we observe.
$E_{p,\text{max}}$ and $E_{p,\text{min}}$ are the maximum energy and minimum energy of the accelerated protons.
Here, we assume that a fraction \( \epsilon_p \) of the dissipated energy in the jet goes into protons, while another fraction \( \epsilon_B \) goes into the magnetic field. Together with the fraction previously mentioned that goes into electrons, the dissipated energy is thus divided into three components in total.
$f_{p\gamma}$ is the pion production efficiency. $f_{\rm cooling}$ represents the fraction of intermediate products, such as $\pi^+$ and $\mu^+$, that have cooled before neutrinos are produced. The detailed derivations of the above can be found in the Appendix \ref{appendix:A}.
\section{The detection prospects of GRB 221009A-\MakeLowercase{like} events} \label{sec:three}

The gamma-ray burst GRB 221009A is often referred to as the “brightest of all time” \citep{Burns_2023,Lesage2023,Williams2023,An2023,Frederiks2023}. 
% stands as the most energetic burst detected to date.
This event was first triggered by the Fermi/Gamma-ray Burst Monitor at 3:16:59 UT on 2022 October 9, with a time-integrated energy flux within the 10–1000 keV range reported as $\left( 2.912 \pm 0.001 \right) \times 10^{-2} \ \text{erg} \ \text{cm}^{-2}$. The peak photon number flux reached $\left( 2385 \pm 3 \right) \ \text{cm}^{-2} \ \text{s}^{-1}$, sustaining this level for $1.024$ s \citep{2022GCN.32642....1L}.
%with a $1.024$s peak photo flux at the level of $\left( 2385 \pm 3 \right) \ \text{cm}^{-2} \ \text{s}^{-1}$, making this the most fluent and intense \GRB\ detected by Fermi-\GBM\ \citep{2022GCN.32642....1L,2023ApJ...946L..26A}.
Using measurements from the Swift observatory, this event was localized at right ascension $\alpha = 288.2645^\circ$ and declination $\delta = +19.7735^\circ$ \citep{2022GCN.32632....1D}, with a host galaxy identified at a redshift of $z = 0.151$. Consequently, its isotropic energy reaches $\sim 10^{55} \ \text{erg}$ \citep{an2023insighthxmtgecamcobservationsbrightestofalltime,Yang_2023,2023ApJ...949L...4L}, 
making it a highly promising candidate to exhibit a neutrino counterpart that can be seen by IceCube, although it was not detected \citep{Abbasi_2023}.
% making it a highly promising candidate to exhibit a neutrino counterpart \textbf{detected by IceCube}, although it was not detected \citep{Abbasi_2023}.

%The \brust\ was located at redshift $z=0.151$\citep{2022GCN.32648....1D}, and have an isotropic radiation energy of
%$\sim  10^{55} \ \text{erg}$\citep{an2023insighthxmtgecamcobservationsbrightestofalltime,Yang_2023,2023ApJ...949L...4L},

\begin{figure}[htbp]
    \centering
    \includegraphics[width=\columnwidth]{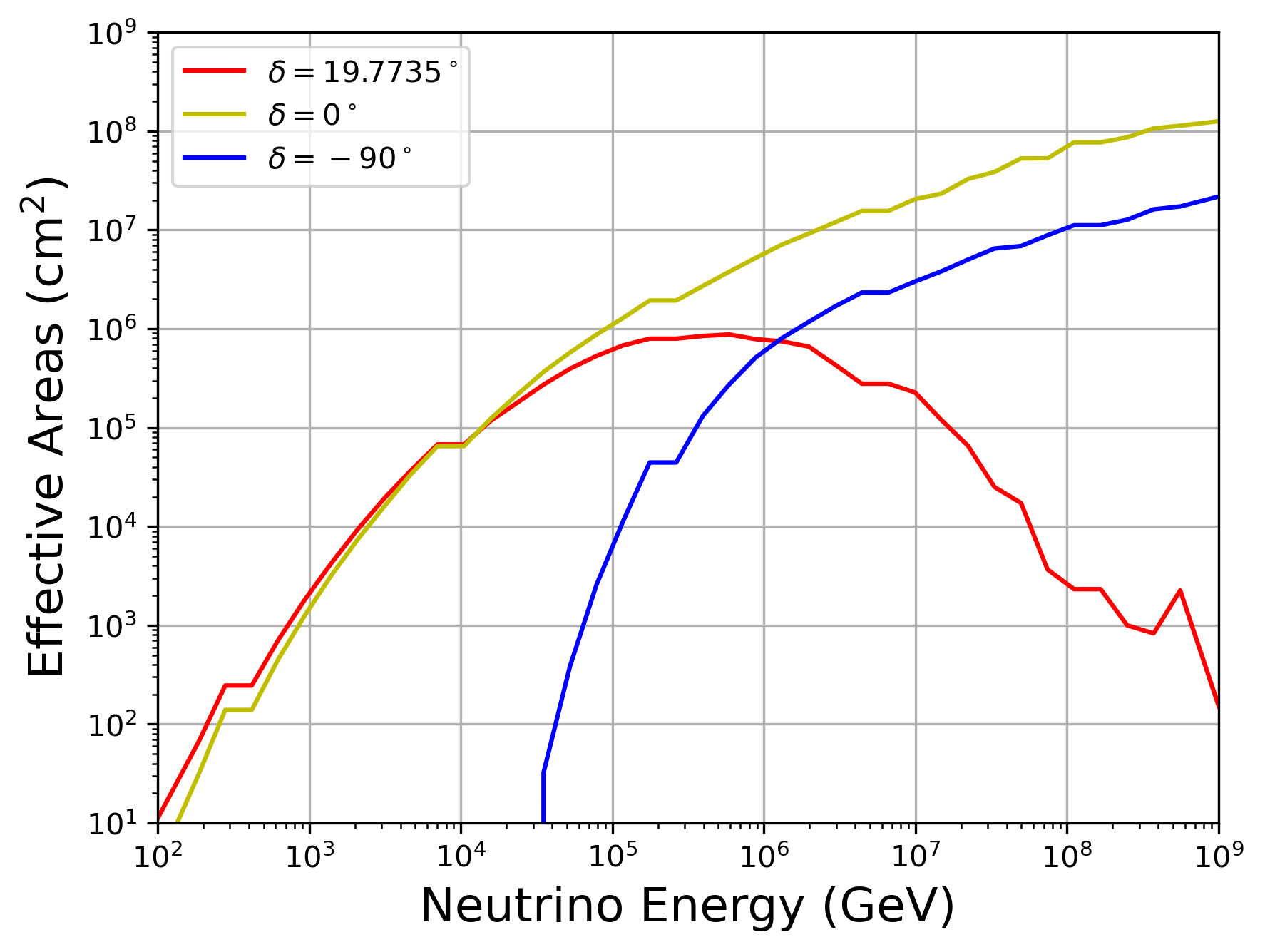}  % 替换为你的图片文件名
    \caption{The effective area as a function of neutrino energy at the declinations of $\delta$ = +19.77°(GRB 221009A), 0°, and -90° for IceCube IC86-II.}
    \label{fig:221009_area}
\end{figure}

Given the predicted neutrino flux, the expected number of events recorded by the neutrino detector can be expressed as
%The number of expected events $N_\nu$ is given by
\citep{icecube_10year}
\begin{equation}
N_{\nu} = \int \text{d}t \int \text{d}\Omega \int_{0}^{\infty} \text{d}E \ A_{\text{eff}} (E, \Omega) \ F_{\nu} (E_{\nu}, \Omega, t) 
\label{eq:neutrino_number}
\end{equation}
where $F_{\nu} = \phi_{\nu}/(E_\nu^2 T)$ is the specific number flux of neutrinos, with $T$ as the duration of observation length.
%which has a dimension ${Gev}^{-1} {cm}^{-2} s^{-1}$.
$A_{\rm eff}$ is the effective areas of the neutrino detector. Here, we use IceCube IC86-II effective area \citep{icecube_10year}, which depends on the neutrino energy and the decl of the source in the sky. The effective areas corresponding to several typical declinations as a function of neutrino energy are shown in Figure \ref{fig:221009_area}. 

Given the expected number of neutrinos to be detected, the probability of actually detecting $N_\nu$ neutrinos is \citep{PhysRevD.110.063004}
\begin{equation}
P_{N_\nu}= 1-\exp(-N_\nu)
\label{eq:neutrino_P}
\end{equation}

\begin{figure}[htbp]
    \centering
    \includegraphics[width=\columnwidth]{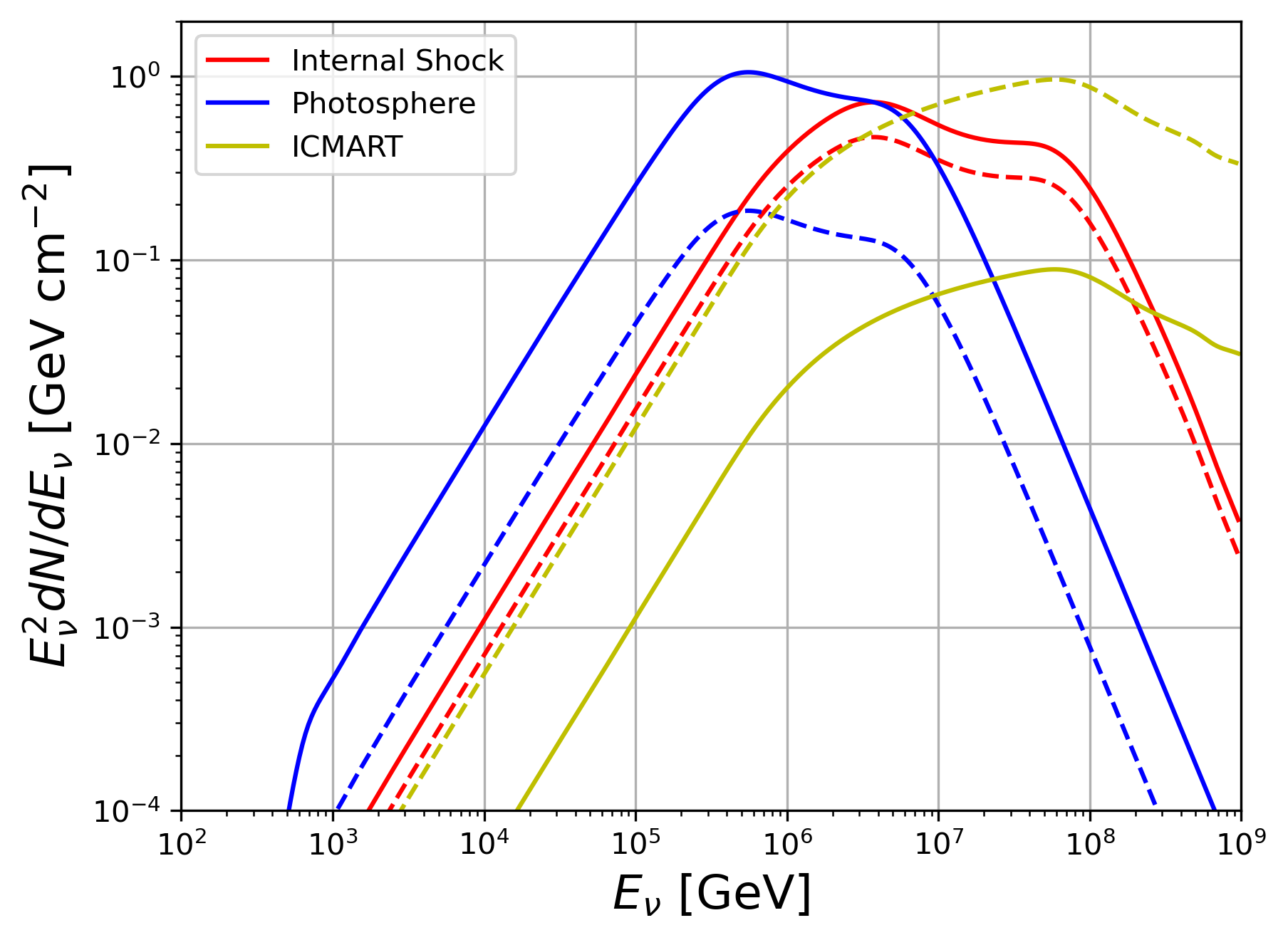}  % 替换为你的图片文件名
    \caption{
    The solid lines represent the predicted neutrino spectrum for GRB 221009A based on the internal shock, dissipative photosphere, and ICMART models, respectively. The indices for the Band function are fitted as $\alpha = 0.97$ and $\beta = 2.37$. The isotropic energy, $E_{\text{GRB}} = 1.15 \times 10^{55} \ \text{erg}$, the isotropic luminosity, $L_{\text{GRB}} = 1.9 \times 10^{52} \ \text{erg} \ \text{s}^{-1}$ and the $\Gamma=300$ are adopted. For all models, $\epsilon_B / \epsilon_e = 1$ and $\epsilon_p / \epsilon_e = 3$ are adopted. For the internal shock model, $\delta t_{\text{min}} = 0.01 \ \text{s}$ is adopted. For ICMART model, $R_{\rm ICMART}=10^{15}$ cm is adopted. The dashed lines represent the $90\%$ confidence-level upper limit of the fluence under nondetection conditions with IceCube, with effective areas of IceCube IC86-II applied.
    %of observing the neutrino with a 90\% probability for each model according to the
    %effective areas of IceCube IC86-II.
    } 
    \label{fig:221009A_fluence}
\end{figure}

The predicted neutrino fluence associated with GRB 221009A from the dissipative photosphere, internal shock, and ICMART models, along with the corresponding $90\%$ upper limit under the nondetection condition with IceCube, is shown in Figure \ref{fig:221009A_fluence}. Here, we adopt $\epsilon_p / \epsilon_e = 3$ and $\epsilon_B / \epsilon_e = 1$ for all models. For the internal shock model, $\delta t_{\text{min}} = 0.01 \, \text{s}$, and for the ICMART model, $R_{\rm ICMART} = 10^{15} $ cm. With the predicted fluence, we can get the corresponding neutrino number to be detected by IceCube, that is, $N_{\rm ph} = 13.0132$,  $N_{\rm IS} = 3.5410$, and $N_{\rm ICMART} = 0.2109$. Thus, for
each model, the detection probability can be calculated as
$P_{\rm ph} = 99.99\% $, $P_{\rm IS} = 97.10\% $, and $P_{\rm ICMART} = 19.02\%$.
We can see that with the current detection capabilities, at the declinations where GRB 221009A appears, the dissipative photosphere and internal shock models have a relatively high probability of detecting neutrinos, while the probability of detecting neutrinos under the ICMART model is relatively low. Therefore, the nondetection fact of high-energy neutrinos associated with GRB 221009A is consistent with the claim that it is driven by the ICMART model, combined with evidence from multiwavelength EM observations \citep{Yang_2023}.
%However, for GRB 221009A, there is a significant chance that it is driven by the ICMART model\citep{Yang_2023}, making the probability of detecting the corresponding neutrinos very slim. 

Despite the unique characteristics of GRB 221009A, primarily its exceptionally bright luminosity, studies like \cite{2023ApJ...949L...4L} suggest that it is a nearby bright GRB with properties similar to those of energetic GRBs.
% Despite GRB 221009A being referred to as a ``once-in-a-millennium'' GRB, it is not particularly unique. \cite{2023ApJ...949L...4L} suggests that it is simply an ordinary nearby GRB with extraordinary observational properties. 
Therefore, in the future, if a GRB 221009A-like event occurs at different sky positions where the neutrino detector can achieve a larger effective area, neutrinos from such events may have chance to be detected.
%similar to 221009A appears in a different declinations and we have better detection capabilities, the detection probability will be shown in 
\begin{figure}[htbp]
    \centering
    \includegraphics[width=\columnwidth]{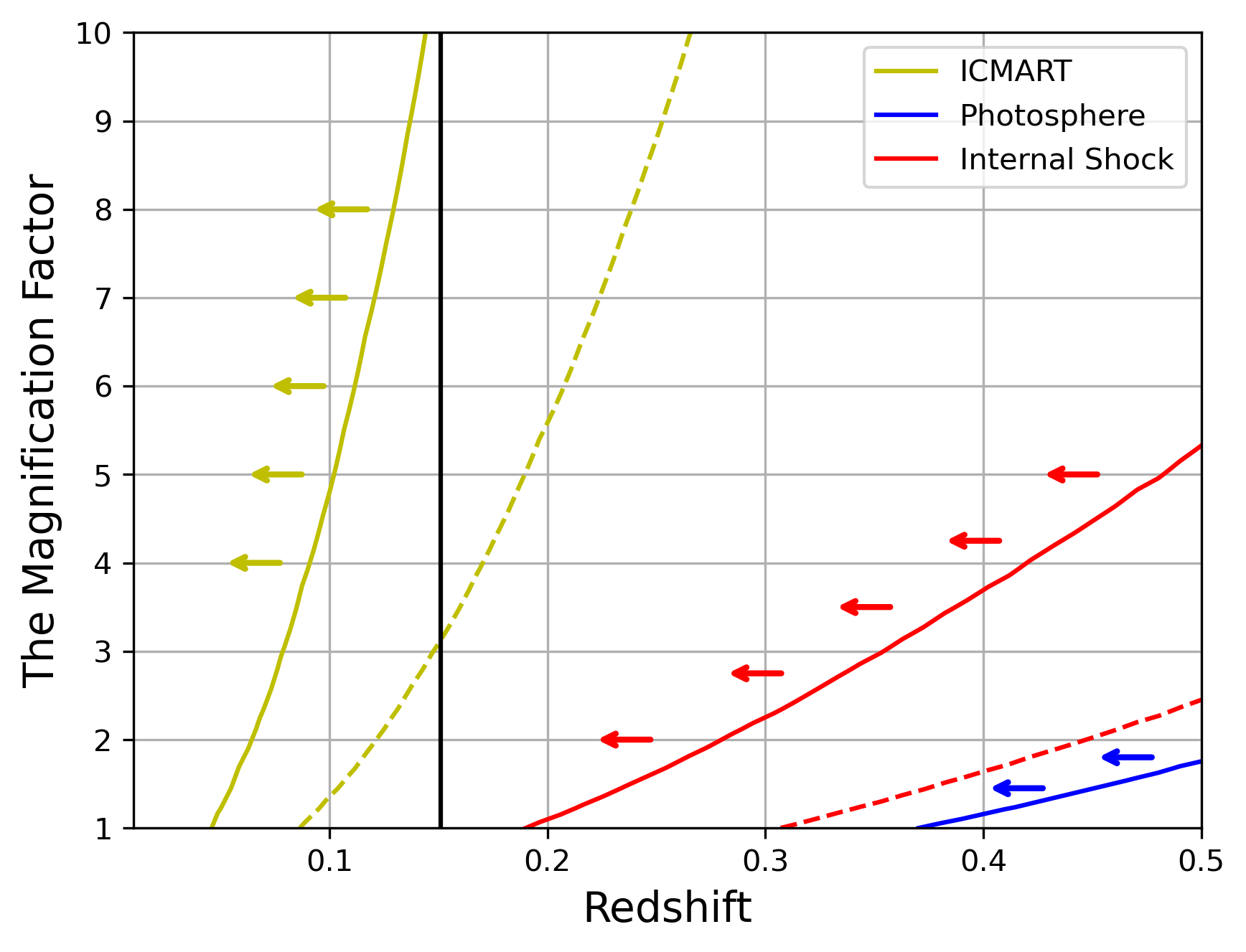}  % 替换为你的图片文件名
    \caption{The prospect to detect a GRB 221009A-like event. The horizontal axis represents the redshift of the event, and the vertical axis represents the magnification factor relative to the effective area of IceCube IC86-II. The black line represents the redshift of the GRB 221009A. The colored solid lines represent the detection limits for different models at the current decl of GRB 221009A, while the dashed lines correspond to scenarios in which GRB 221009A is located near the equator, where the effective area is maximized. The effective area is calculated by integrating over the neutrino energy range from $10^2$GeV to $10^9$ GeV. The arrows indicate the parameter spaces in which the source is detectable.} 
    \label{fig:221009_large}
\end{figure}
It should be noted that, according to the conclusions of \cite{Ai_2023}, for GRB 221009A, only if its internal shock originates from a region with a very large dissipation radius (a large variability timescale or a very large bulk Lorentz factor) can it match our expectation of not detecting neutrinos. \cite{Rudolph_2023} presented a more sophysticated internal shock model, where the variability timescale is around $1$ s, and the final energy dissipation occurs at a radius of approximately 
$ 10^{16}\sim 10^{17}$ cm, which is even larger than that of the ICMART model. In this discussion, we adopt a conservative approach and assume that the minimum variability timescale is the classical $0.01$ s \citep{PhysRevD.83.067303,H_mmer_2012,Aartsen_2017}. As shown in Figure \ref{fig:221009_large}, in the context of the dissipative photosphere model, placing GRB 221009A at a redshift of 0.37 would still allow its generated neutrinos to be detectable by the current IceCube detector. With a less than twofold increase in the detector's sensitivity, the neutrinos produced by GRB 221009A would remain observable even at a redshift of 0.5. In the context of the internal shock model, neutrinos from GRB 221009A can be detected within a redshift of $0.19$ without any increase in the detector's effective area. A fivefold increase in the detector's effective area would enable detection of neutrinos, provided the event occurs within a redshift of 0.5. In the case of the ICMART model, if the event occurs at the same redshift as GRB 221009A (z = 0.15), a tenfold increase in the detector's effective area would be required to have a chance of detecting the neutrinos. It is worth noting that the above calculations are based on the effective area corresponding to the true decl angle of GRB 221009A. If a future event occurs at a decl where the detector's effective area is maximized (i.e., near the equator), the detection rate would increase significantly. This maximized effective area is obtained by integrating over the predicted neutrino energy range from $10^2$ GeV to $10^9$ GeV. In that case, only about $3$ times the increase of the detector's sensitivity would be needed to detect the neutrinos from GRB 221009A-like bursts, at $z=0.15$ for the ICMART model or at $z=0.5$ for the internal shock model. The above conclusions are based on a 90\% detection probability. Please note that the above discussion is based on the parameters used in Figure \ref{fig:221009A_fluence}. The flux of neutrinos is influenced by these parameters, which may lead to potential uncertainties.

%\textcolor{red}{(neutrino energy range of $ 10^2 $ Gev to $ 10^9 $ Gev)}, \textcolor{red}{specifically within the declination ranges of  $-2.29^\circ$ to $0^\circ$ for the ICMART model and $0^\circ$ to $2.29^\circ$ for the internal shock model }

%\section{THE STACKED NEUTRINOS FROM 2019-2023 LONG \GRB} \label{sec:four}

\section{STACKED NEUTRINOS FROM LONG GRBs} \label{sec:four}

%Through our analysis of GRB 221009A in Section \ref{sec:three}, we found that relying solely on a single source, even one as bright as GRB 221009A, makes it quite difficult to observe neutrinos from the GRB, especially given the current detection capabilities of high-energy neutrino detectors.

If we are not "lucky" enough to encounter an event similar to GRB 221009A, we will have to rely on the stacked detection approach. Although the chance of detecting high-energy neutrinos associated with a single ``normal" GRB event is low, the accumulation of GRB events increases the probability of detecting a high-energy neutrino associated with a GRB, which could eventually reach a considerably high level.

%We know that we can observe roughly GRB per day, which means we have a chance to detect neutrinos every day. However, each opportunity has a probability as low as one in a thousand or even one in ten thousand. But with the accumulation of time and gamma-ray bursts, the probability of detecting neutrinos from a GRB will increase significantly.

Here, we use data from GRBweb\footnote{For detailed data, see \url{https://user-web.icecube.wisc.edu/~grbweb_public}.}, an IceCube project that collects data from different telescopes, such as GBM \citep{Hurley_2013,vonKienlin_2020}, LAT \citep{Ajello_2019}, Swift \citep{Lien_2016} and others. From 2019 to 2023, a total of 1503 gamma-ray bursts were recorded. We select those with fluence records and $T_{90}>2$ s for our calculations. 
% \textbf{These include both low-luminosity (LL) GRBs and high-luminosity (HL) GRBs. Some studies suggest that LL GRBs, due to their higher event rate, contribute significantly, or even more, to the total diffuse neutrino flux compared to HL GRBs \citep{Murase_2006_ll,GUPTA2007386}. However, in this work, we only consider the actual observational scenario, where a single neutrino event is linked to a specific GRB event.} 
If a source did not have a redshift measurement, we assigned a redshift value of $2.15$\citep{Aartsen_2017}. We adopt $ E_{\text{break}} = 200 $ kev, $\alpha=1$ and $\beta=2$ for all GRBs. For the bulk Lorentz factor ($\Gamma$) of GRBs, we derive it using the empirical relationship between the bulk Lorentz factor and the luminosity of the GRB \citep{Liang_2010,Lü_2012,zhang_2013}, which is given by 
\begin{equation}
    \Gamma \sim  250  L_{\text{iso},52}^{0.30}
\end{equation}
%\textcolor{red}{where $A_X=A/10^X$}.
For simplicity, the uncertainties connected with this relation, which may result from the orientation of the jets, are not considered.
We exclude GRB 210518A and GRB 230614C, even though their fluences were well recorded in GRBweb. This is because, in the absence of redshift measurements, assuming a redshift of 2.15 for these sources would result in an unreasonably high luminosity. As a result, they would produce a number of neutrinos comparable to those of the remaining $\sim 1000$ gamma-ray bursts. In addition, GRB 221009A is also excluded from the samples.

\begin{figure}[htbp]
    \centering
    \includegraphics[width=\columnwidth]{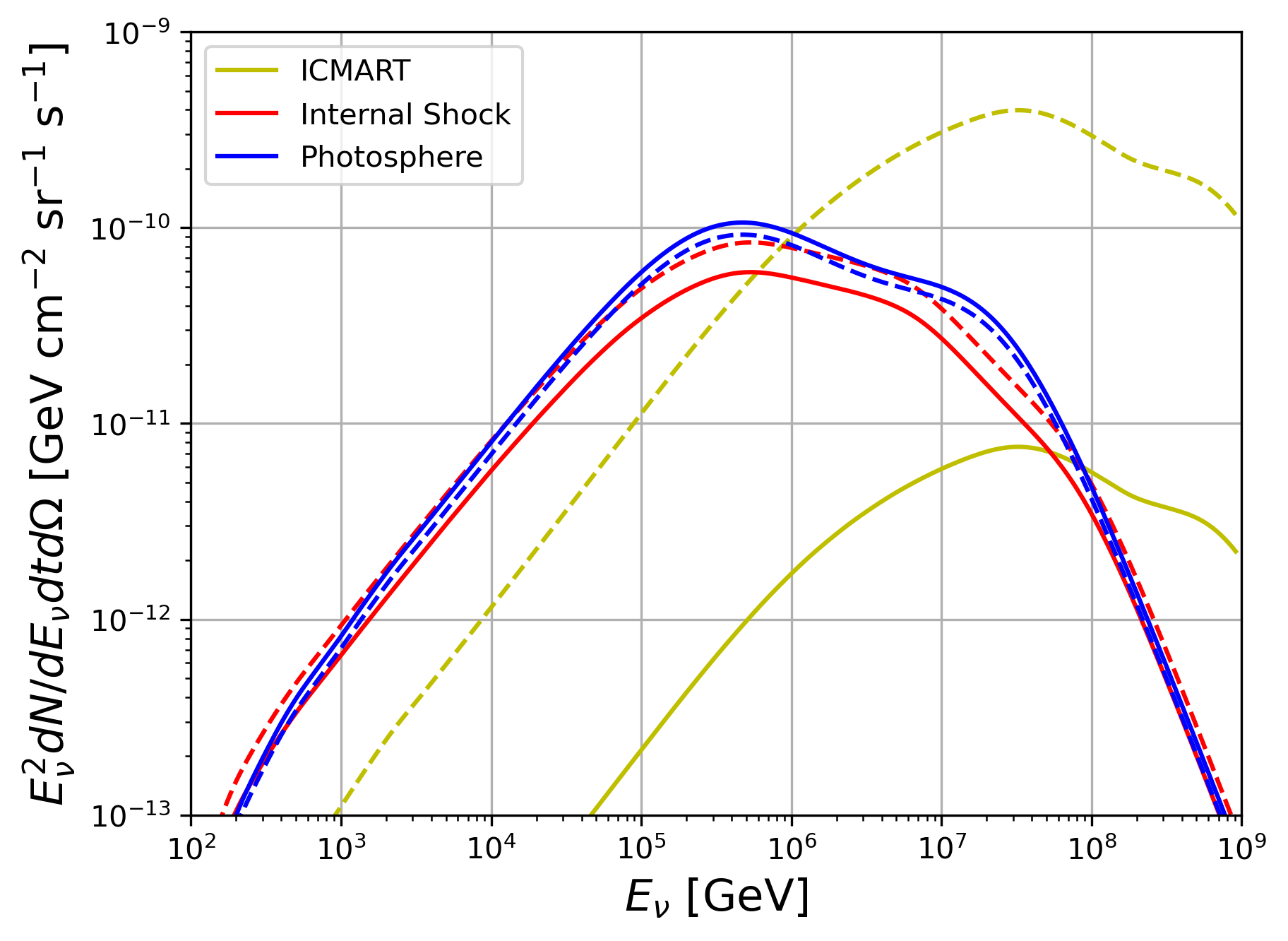}  % 替换为你的图片文件名
    \caption{The solid lines are the predicted neutrino flux for the internal shock, dissipative photosphere, and ICMART models. The dashed lines represent the upper limit 
    of observing the neutrino with a 90\% probability for each model according to the effective areas of IceCube IC86-II. $\epsilon_p/\epsilon_e = 3$, $\epsilon_B/\epsilon_e = 1$, $\delta t_{\rm min}=0.01$ s, and $R_{\rm ICMART}=10^{15}$ cm are adopted.}
    \label{fig:22019_2023_model}
\end{figure}

% \subsection{Current observational status} \label{sec:four_one}
The average stacked neutrino fluxes produced by all-sky GRB events from 2019 to 2023, assuming the dissipative photosphere, internal shock, and ICMART models, are shown in Figure \ref{fig:22019_2023_model}. Here, we also adopt $\epsilon_B / \epsilon_e = 1$ and $\epsilon_p / \epsilon_e = 3$ for all models. For the internal shock model, $\delta t_{\text{min}} = 0.01\ \text{s}$ is adopted. For the ICMART model, $R_{\rm ICMART}=10^{15}$ cm is adopted. Based on these fluxes, we calculate the expected number of neutrinos detected by IceCube for the three models as $N_{\rm ph} \approx 2.65$, $N_{\rm IS} \approx 1.62$, and $N_{\rm ICMART} \approx 0.0439$, corresponding to the detection probabilities of $P_{\rm ph} \approx 92.90\%$, $P_{\rm IS} \approx 80.19\%$, and $P_{\rm ICMART} \approx 4.293\%$, respectively. We can see that for the ICMART model, there is still a significant gap to reach a 90\% detection probability, whereas for the dissipative photosphere and internal shock models, the probability of detecting neutrinos is approximately 90\%. Please note that we have assumed uniform benchmark microphysical parameters for all GRBs and employed an approximate relation to determine the bulk Lorentz factor of the jet. Various parameter settings might lead to different outcomes.

\begin{figure}[htbp]
    \centering
    \includegraphics[width=\columnwidth]{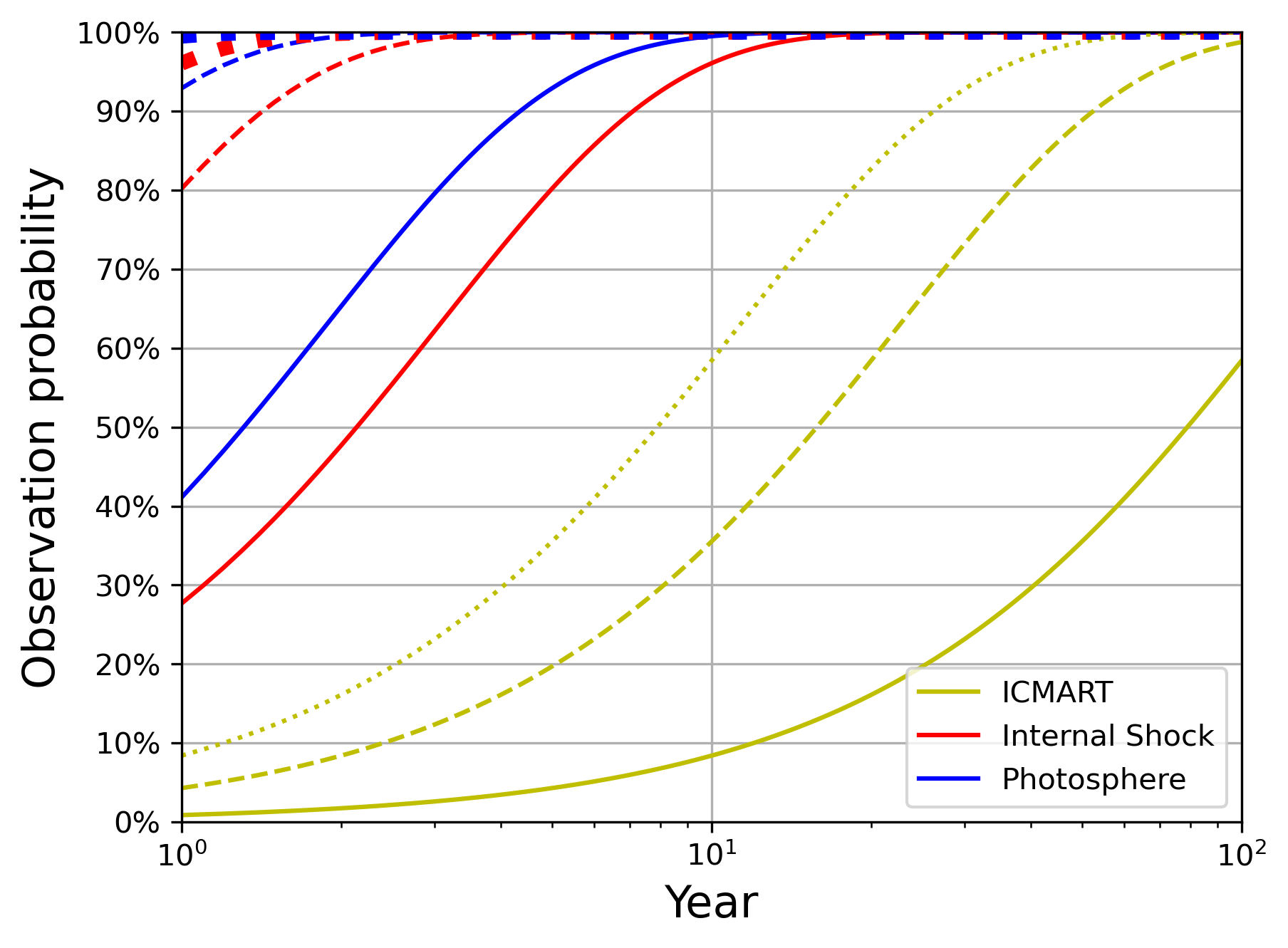} 
    \caption{For different models, the detection probability of neutrinos varies with the accumulation time. The solid line represents the current effective area of IceCube, while the dashed and dotted lines represent the effective area expanded by a factor of 5 and 10, respectively. The dotted lines corresponds to the dissipative photosphere, and the internal shock models have been bolded for better visibility. $\epsilon_p/\epsilon_e = 3$, $\epsilon_B/\epsilon_e = 1$, $\delta t_{\rm min}=0.01$ s, and $R_{\rm ICMART}=10^{15}~{\rm cm}$ are adopted.}
    \label{fig:22019_2023_time}
\end{figure}

We assume that GRBs will continue to be observed in the coming years at the same detection rate as during the period from 2019 to 2023. Using the same parameters as those applied in the models shown in Figure \ref{fig:22019_2023_model}, we calculate the evolution of the probability of detecting neutrinos from GRBs over the detector's operational time.
%We select a specific set of parameters and consider different accumulation times and magnification factors of detector effective areas. 
The results are shown in Figure \ref{fig:22019_2023_time}. Assuming the detector's effective area remains unchanged, the dissipative photosphere model and the internal shock model require 4.35 yr and 7.11 yr, respectively, to achieve a $90\%$ detection probability. In contrast, the ICMART model can only reach a detection probability of $58\%$, even with an accumulation time of 100 yr.

With next-generation neutrino detectors, the increase in the effective area will significantly enhance the probability of jointly detecting GRBs and neutrinos. If the detector’s effective area is increased fivefold
relative to IceCube IC86-II, the dissipative photosphere and internal shock models would require only 1 yr of accumulation to achieve detection probabilities of 80\% and 93\%, respectively. In contrast, the ICMART model would require 52 yr of accumulation to reach a 90\% detection probability. With a tenfold expansion of the detector’s effective area, the dissipative photosphere and internal shock models would quickly approach a 100\% detection probability. However, even with 10 years of accumulation, the ICMART model would only achieve a 58\% detection probability.

\begin{figure}[htbp]
    \centering

    \begin{minipage}{0.45\textwidth}
        \centering
        \parbox{\textwidth}{\centering Dissipative photosphere} % 使用 \parbox 显示子图标题
        \includegraphics[width=\textwidth]{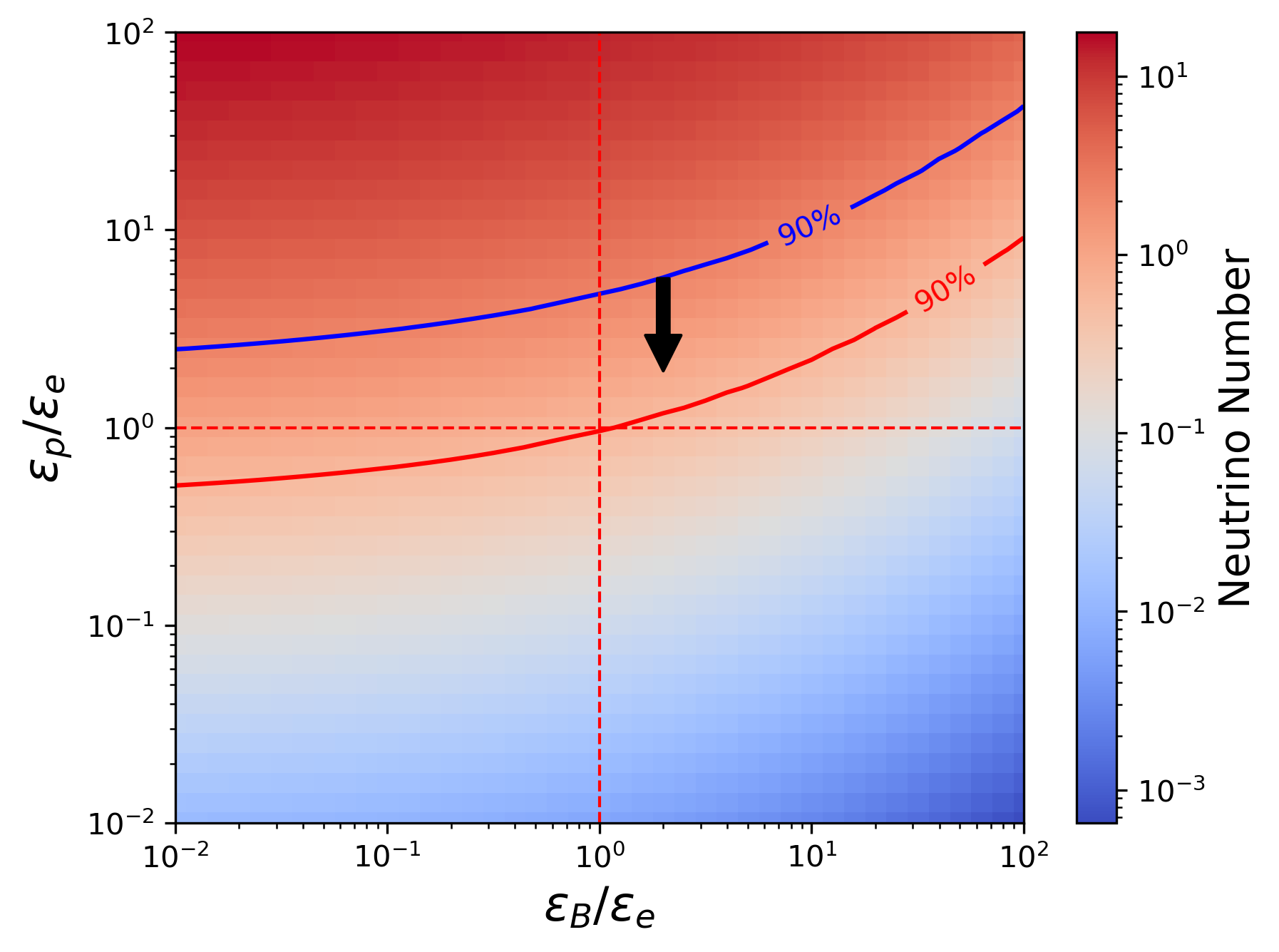} % 替换为图片路径
        
    \end{minipage}

    % 子图 a
    \begin{minipage}{0.45\textwidth} % 设置子图宽度为 45%
        \centering
        \parbox{\textwidth}{\centering Internal Shock} % 使用 \parbox 显示子图标题
        \includegraphics[width=\textwidth]{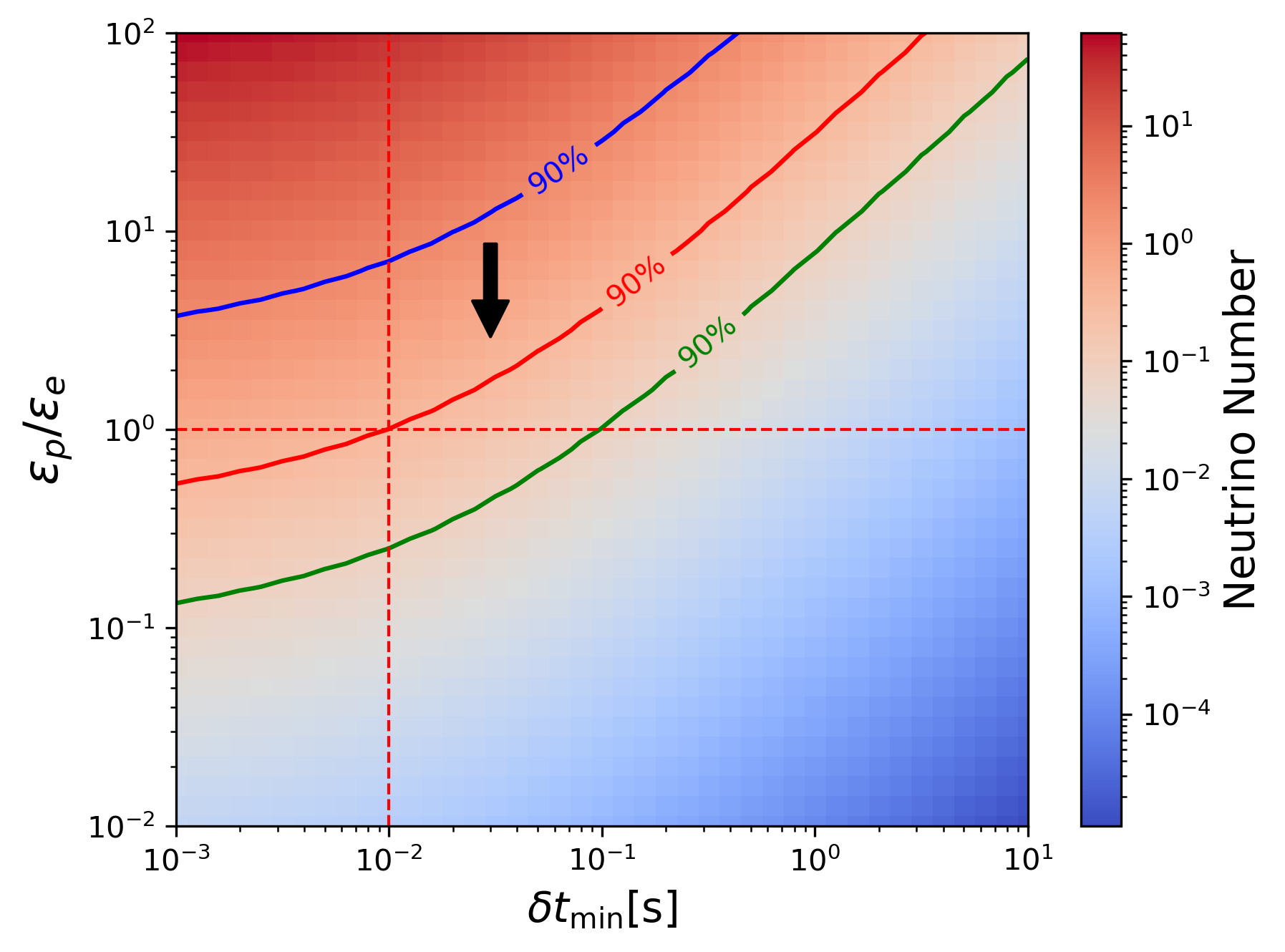} % 替换为图片路径
        
    \end{minipage}
    \hspace{0.05\textwidth} % 设置图像之间的间距
    % 子图 b

    \begin{minipage}{0.45\textwidth}
        \centering
        \parbox{\textwidth}{\centering ICMART} % 使用 \parbox 显示子图标题
        \includegraphics[width=\textwidth]{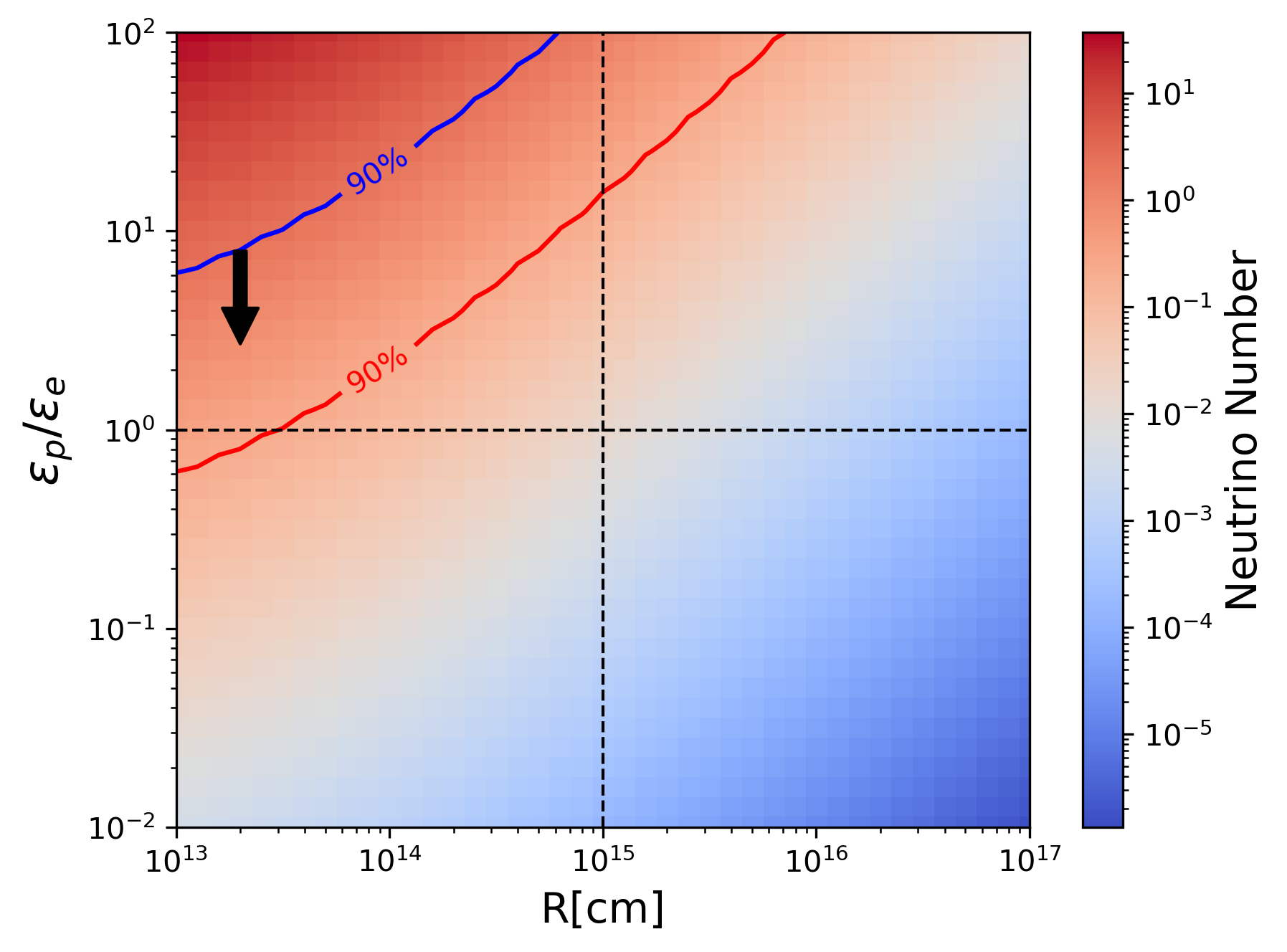} % 替换为图片路径
        
    \end{minipage}

    \caption{The solid lines represent the upper limits for which there is a 90\% probability of detection. The parameter space closer to the lower right corner is more tolerable. For all three models, the blue line corresponds to the current effective area of IceCube IC86-II. For the dissipative photosphere models, red line represent the current effective area expanded 4 times relative to IceCube IC86-II. For the internal shock models, red and green line represents the effective area expanded 5.5 and 19 times relative to IceCube IC86-II, respectively.For the ICMART model, red line represents the effective area expanded 10 times relative to IceCube IC86-II.}
    \label{fig:2019_2023_large}
\end{figure}

On the other hand, even if the neutrino counterparts of GRBs remain undetected, much more stringent constraints can be placed on the free parameters of different GRB models. If the parameters are constrained to an unacceptable range, it can be concluded that the corresponding GRB model can be ruled out. Inspired by observations of GRBs and their afterglows, we assume reasonable parameters to be $\epsilon_p / \epsilon_e > 1$, $\epsilon_B/\epsilon_e < 1$ \citep{Gao_2015}. And we still adopt $\delta t_{\rm min} =0.01$ s
% \citep{MacLachlan_2013,bhat2013variabilitytimescaleslong} 
and $R_{\rm ICMART} = 10^{15}$ cm. In the future,
assuming the events accumulated over 5 yr, similar to those from 2019 to 2023, using the nondetection results from an enhanced neutrino detector and the parameter space shown in Figure \ref{fig:2019_2023_large}, one may conclude that:

\begin{itemize}
    \item For the dissipative photosphere model, if the effective area has been increased by a factor of 4 relative to IceCube IC86-II, then for $\epsilon_p/\epsilon_e > 1$, $\epsilon_B/\epsilon_e > 1.18$ is required, suggesting that this model is not generally applicable to GRBs.
    \item  For the internal shock model, if the effective area has been increased by a factor of $5.5$ relative to IceCube IC86-II, then for $\delta t_{\rm min} = 0.01$ s, $\epsilon_p/\epsilon_e$ must be less than approximately $0.96$. This implies that the internal shock model with $\delta t_{min} = 0.01 $s is not generally applicable to GRBs.
    In addition to the conservative case, some also suggest that the minimum variability timescale could be 0.1 s \citep{zhang_2013}. We have also considered this scenario and find that if the detector's effective area is increased by a factor of 19, and neutrinos are still not detected, then for $\delta t_{\rm min}=0.1$ s, $\epsilon_p/\epsilon_e$ must be less than 1. Therefore, to rule out this scenario, the detector's effective area would need to be increased by a factor of 19.
    \item For the ICMART model, if the effective area has been increased by a factor of $10$ relative to IceCube IC86-II, then for $R_{\rm ICMART} = 10^{15}~{\rm cm}$, $\epsilon_p / \epsilon_e < 15$ is required. This is consistent with the theoretical description, so we cannot impose strong constraints on the ICMART model.If we want to constrain $ \epsilon_p / \epsilon_e $ to below 1 without detecting neutrinos for $ R_{\rm ICMART} = 10^{15}$ cm, the detector's effective area would need to be increased by a factor of 150. This is far beyond the detection capabilities of both our current and near-future detectors.
\end{itemize}

\section{Predictions with upcoming neutrino detectors}

\begin{figure}[htbp]
    \centering
    \includegraphics[width=\columnwidth]{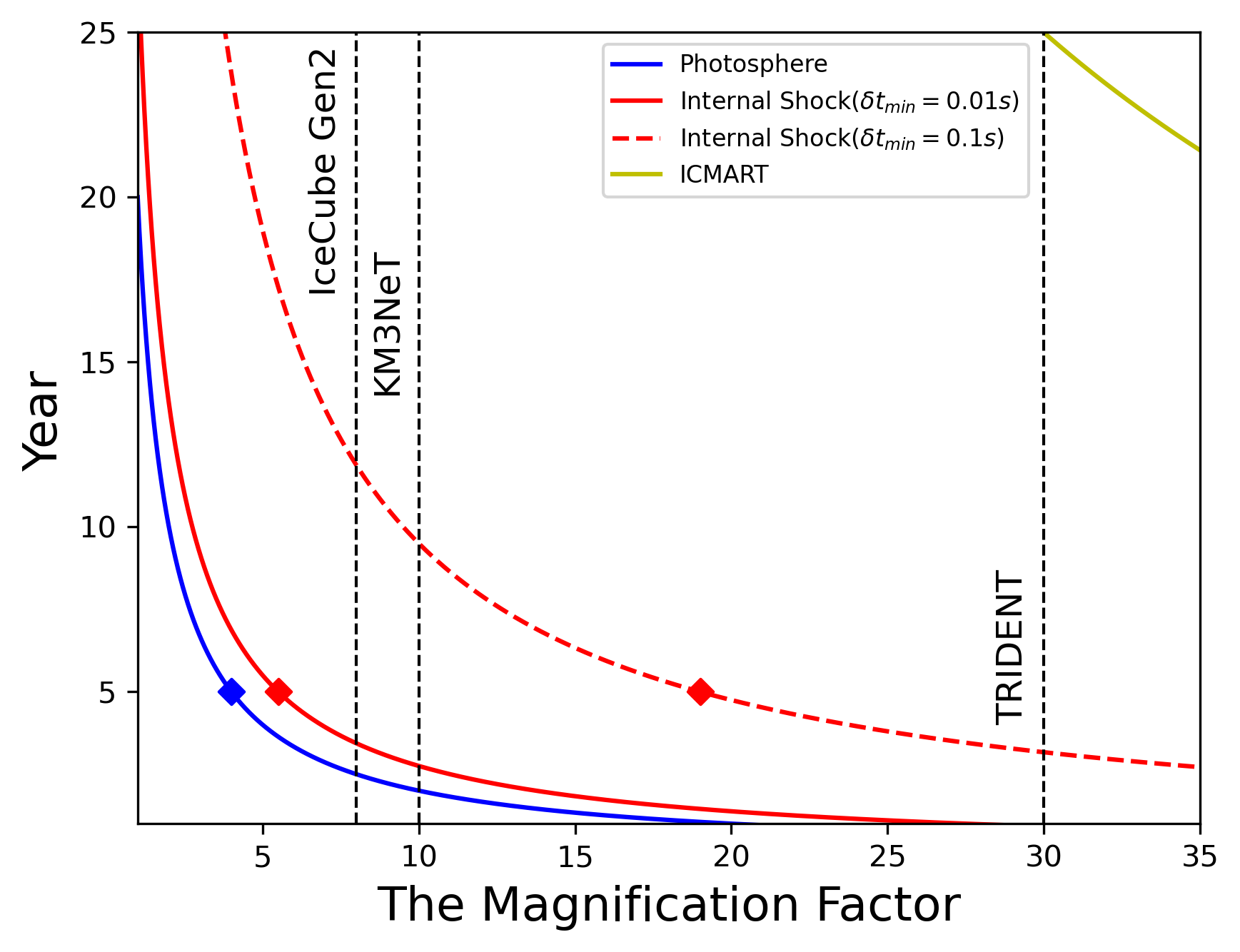}  % 替换为你的图片文件名
    \caption{The required magnification factors and data accumulation periods for future neutrino detectors to effectively rule out different GRB prompt emission models are shown. The upper-right region of each line represents the required parameter space. The three dots indicate the required magnification factors of the detectors' effective areas, relative to IceCube IC86-II, to exclude the dissipative photosphere model and the internal shock model ($\delta t_{\rm min}=0.01s$ or $\delta t_{\rm min}=0.1s$) under a five-year observation. The three vertical black dashed lines represent the magnification factors of IceCube Gen2, KM3NeT, and TRIDENT.}
    \label{fig:future decs}
\end{figure}

A number of innovative neutrino detectors have been proposed, and several are presently being constructed. With their proposed sensitivity curves, one can estimate the magnification factors with respect to IceCube IC86-II. Here we assume the neutrino spectrum follows \( E^{-2} \). By integrating over the energy range from \( 10^2 \) GeV to \( 10^9 \) GeV using this spectrum, we obtain IceCube IC86-II’s sensitivity at different declinations. Since IceCube's detection capability is highly sensitive to decl, we take the average sensitivity across different values of the sine of the decl and compare this average with the sensitivities of other detectors, which are shown in the corresponding references. We find that the magnification factors are $\sim 8$ for IceCube Gen2 \citep{omeliukh2021optimizationopticalarraygeometry}, $\sim 10$ for KM3NeT \citep{AIELLO2019100}, and $\sim 30$ for TRIDENT \citep{ye2024multicubickilometreneutrinotelescopewestern}. Comparing these magnification factors with the requirements shown in Figure \ref{fig:221009_large}, we find that all future neutrino detectors can achieve an ideal detection prospect for a single source resembling GRB 221009A, provided the source occurs at the same redshift. IceCube Gen2, with a relatively low magnification factor, may not be fully located within the detectable parameter space for all GRB prompt emission models.
% Comparing these magnification factors with the requirements shown in Figure \ref{fig:221009_large}, we find that for a single source resembling GRB 221009A, all future neutrino detectors can achieve an ideal detection prospect if the source occurs at the same redshift, although IceCube Gen2 may not be fully located within the detectable parameter space for all GRB prompt emission models.
For the stacked neutrinos from GRBs, based on the general results shown in Section \ref{sec:four}, we explicitly outline the conditions required to support or effectively rule out different GRB models in Figure \ref{fig:future decs}, if no neutrino-GRB association is confirmed, with the proposed neutrino detectors' effective areas marked. All three proposed neutrino detectors have the capability to rule out the dissipative photosphere model and the internal shock model with $\delta t_{\rm min}=0.01s$, over a 5 yr period of operation and can very likely achieve this progress within 2 to 3 yr. For the internal shock model with $\delta t_{\rm min}=0.1s$, only TRIDENT has the capability to rule it out within 5 yr (specifically in 3.17 yr). For the ICMART model, the required magnification factor is far beyond the capabilities of both current and near-future detectors. However, \cite{Agostini_2020} proposed that multiple neutrino detectors can be combined into a network, which would significantly enhance detection capabilities and potentially meet the requirement to rule out the ICMART model.

\section{Conclusions and Discussions} \label{sec:five}

GRBs are potential sources of high-energy neutrinos. However, despite extensive studies, including the exceptionally bright GRB 221009A and over a decade of cumulative neutrino searches, no definitive association has been confirmed.

The lack of neutrino detections provides meaningful insights into models of GRB prompt emission. Stringent constraints have been placed on the physical parameters of the dissipative photosphere and internal shock models, while the parameter space for the ICMART model remains broad.

In this work, we first calculate the neutrinos produced in a GRB 221009A-like event under the dissipative photosphere, internal shock, and ICMART models, respectively. Our calculations indicate that, under typical parameters, if GRB 221009A originated from either the dissipative photosphere model or the internal shock model, its neutrinos should have already been detected. 

Thus, the lack of neutrinos associated with GRB 221009A is consistent with implications from multiband EM observations suggesting that a magnetically dominated jet was launched. With future enhanced neutrino detectors, if the effective area is approximately $10$ times larger than that of IceCube IC86-II, we would be able to detect neutrinos from such a GRB event that has the same redshift with GRB 221009A, even if produced under the ICMART model. If we are particularly lucky, and the event occurs at a decl corresponding to the effective area of maximum detector efficiency, then increasing the effective area by a factor of $3$ would be sufficient to detect the neutrinos it produces.

We then calculated the cumulative neutrino flux from stacked GRBs and analyzed 1142 sources from 2019 to 2023. 
% Our findings indicate that within a significant parameter space, if all GRBs followed either the dissipative photosphere or the internal shock model, the resulting neutrino production would exceed current observational limits. This suggests that these models likely account for only a subset of all GRBs. In contrast, the ICMART model remains consistent with observational limits in most of the parameter space.
We considered a scenario where future detectors with an increased effective area observe these 1142 sources over a 5 yr period. If the effective area is increased $4$ times relative to IceCube IC86-II and no neutrinos are detected, the general applicability of the dissipative photosphere model would be strongly questioned. If expanded $5.5$ times, the same issue appears in the internal shock model. 
For the ICMART model, even if the detector's effective area is increased by a factor of $10$ and no associated neutrinos are detected, the model can still survive.
We also compare the above results with the capabilities of upcoming neutrino detectors and find that they can probe the dissipative photosphere model and the internal shock model. However, a significant gap still remains for the ICMART model.

% Here are three cautions
Please note the underlying conditions required to derive these results: (1) For the internal shock model, we assume the minimum variability timescale of the GRB light curve is $\delta t_{\rm min} \sim 0.01$ s, which is a classical theoretical value. However, if $\delta t_{\rm min}$ is much greater (like $0.1$ s inferred from some of observed minimum variability timescale \citep{golkhou2015,camisasca2023}), the internal shock model should have a radiation radius comparable to that of the ICMART model \citep{Rudolph_2023}, making neutrino production in the internal shock model also very inefficient. (2) When we rule out models using stacked GRB observations and predicted effective areas from future neutrino detectors, 
we mean ruling out the possibility that a single model applies to all GRBs. In fact, there may be multiple channels responsible for producing GRBs. (3) Our discussion is valid only in the ``one-zone" scenario, where protons are accelerated in the same region where the gamma-ray photons are emitted.

Studies predict that low-luminosity GRBs might be more efficient generators of high-energy neutrinos \citep{Murase_2006_ll,GUPTA2007386}. Similarly, short GRBs with relatively lower bulk Lorentz factors in their jets could also be potential sources of high-energy neutrinos \citep{Rudolph2024}. Currently operating powerful gamma-ray and X-ray detectors could detect more of these relatively faint events, thereby providing better constraints on GRB models.

%\noindent author year, title, version, publisher, prefix:identifier\\

%\citet{2015ApJ...805...23C} provides a example of how the citation in the
%article references the external code at
%\doi{10.5281/zenodo.15991}.  Unfortunately, bibtex does
%not have specific bibtex entries for these types of references so the
%``@misc'' type should be used.  The Repository tutorial explains how to
%code the ``@misc'' type correctly.  The most recent aasjournal.bst file,
%available with \aastex\ v6, will output bibtex ``@misc'' type properly.

%% IMPORTANT! The old "\acknowledgment" command has be depreciated. It was
%% not robust enough to handle our new dual anonymous review requirements and
%% thus been replaced with the acknowledgment environment. If you try to 
%% compile with \acknowledgment you will get an error print to the screen
%% and in the compiled pdf.
%% 
%% Also note that the akcnowlodgment environment does not support long amounts of text. If you have a lot of people and institutions to acknowledge, do not use this command. Instead, create a new \section{Acknowledgments}.

\section*{acknowledgments}
% We thank Irene Tamborra for useful comments.
% This work is supported by the National Natural Science Foundation of China (Projects 12373040,12021003)
% and the Fundamental Research Funds for the Central Universities. S.A. has received support from the Carlsberg Foundation (CF18-0183, PI: I. Tamborra).

\begin{acknowledgments}
We thank Irene Tamborra for useful comments.
This work is supported by the National Natural Science Foundation of China (Projects 12373040,12021003)
and the Fundamental Research Funds for the Central Universities. S.A. has received support from the Carlsberg Foundation (CF18-0183, PI: I. Tamborra).
\end{acknowledgments}

%% To help institutions obtain information on the effectiveness of their 
%% telescopes the AAS Journals has created a group of keywords for telescope 
%% facilities.
%
%% Following the acknowledgments section, use the following syntax and the
%% \facility{} or \facilities{} macros to list the keywords of facilities used 
%% in the research for the paper.  Each keyword is check against the master 
%% list during copy editing.  Individual instruments can be provided in 
%% parentheses, after the keyword, but they are not verified.

% \vspace{5mm}
% \facilities{...}

%% Similar to \facility{}, there is the optional \software command to allow 
%% authors a place to specify which programs were used during the creation of 
%% the manuscript. Authors should list each code and include either a
%% citation or url to the code inside ()s when available.

% \software{... 
%           }

%% Appendix material should be preceded with a single \appendix command.
%% There should be a \section command for each appendix. Mark appendix
%% subsections with the same markup you use in the main body of the paper.

%% Each Appendix (indicated with \section) will be lettered A, B, C, etc.
%% The equation counter will reset when it encounters the \appendix
%% command and will number appendix equations (A1), (A2), etc. The
%% Figure and Table counter will not reset.

\appendix

\section{Formulae for calculating neutrino flux}
\label{appendix:A}

Equation \ref{eq:neutrino_flux} shows the general formula for calculating the flux of neutrinos from a single GRB source. Here, we describe the details of each term in the equation.
The maximum proton energy $E_{\rm p,max}$ could be estimated by equaling the dynamical timescale $t' \approx R/\Gamma c$ with the accelerating timescale of protons $t'_{\rm acc} \approx E'_p/\left(eB'c\right) $, where $\Gamma$ is bulk motion Lorentz factor of the GRB jet. The ``$\prime$" denotes physical quantities defined in the jet’s comoving frame, and hereafter. Hence, we have
\begin{equation}
E_{p, \text{max}} \leq 7.59 \times 10^{11} \text{ GeV} \left(\frac{\epsilon_B}{\epsilon_e}\right)^{1/2} L_{\text{GRB},52}^{1/2} \Gamma_{2.5}^{-1}
% $}
\end{equation}
where $L_{\text{GRB}}$ is the isotropic luminosity of the GRB. A lower limit for minimum proton energy can be set as  
\begin{eqnarray}
    E_{p,\text{min}} > \Gamma m_p c^2 = 2.56 \times 10^2 \text{ GeV }  \Gamma_{2.5}.
\end{eqnarray}

% \begin{equation}
% E_{p,\text{min}} > \Gamma m_p c^2 = 2.56 \times 10^2   \Gamma_{2.5} \text{ GeV}  
% \end{equation}
% The proton spectrum can be expressed as a power law that ${dN_p}/{dE_p} \propto E_p^{-s}$ with s $ = 2$ adopted \citep{kimura2022}. Specifically, it is written as \begin{equation}
% \begin{aligned}
% \frac{dN_p}{dE_p}
% &= \frac{\left(\epsilon_p/\epsilon_e\right) E_{\text{GRB}} E_p^{-2}}{\ln\left(E_{p,\text{max}}/E_{p,\text{min}}\right)},
% \end{aligned}
% \end{equation}
% where $E_{\text{GRB}}$ is the isotropic energy of the GRB. 
% We assume that a fraction $\epsilon_e$ of the dissipated energy is transferred to electrons and fully radiated as gamma rays, a fraction $\epsilon_p$ is transferred to protons, and a fraction $\epsilon_B$ goes into the random magnetic field.

\begin{figure}[htbp]
    \centering
    \includegraphics[width=0.7\textwidth]{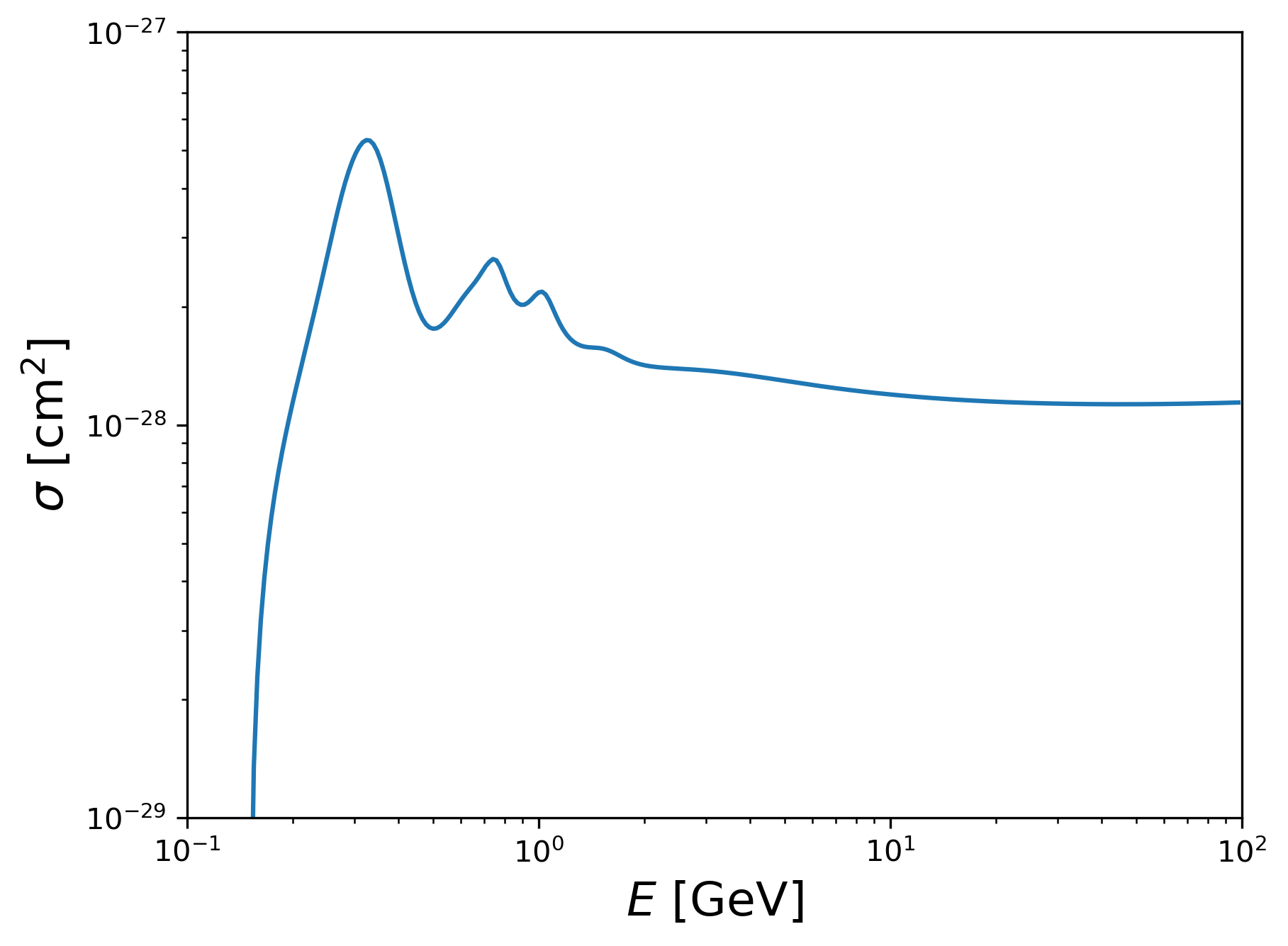}  % 替换为你的图片文件名
    \caption{The horizontal axis represents the photon energy in the rest frame of the proton, and the vertical axis represents the cross section for the $p\gamma $ interaction.
    % The points represent experimental values, while the solid line represents calculated values based on the fitted model.
    }
    \label{fig:sigama}
\end{figure}

Then, we calculate the $p\gamma$ interaction efficiency, which is in principle determined by the production timescale $t_{p\gamma}$ and the dynamical timescale $t_{\rm dyn}$ of the radiation zone in the GRB jet. 
The pion dynamical timescale can be estimated as
$t_{\rm dyn} \approx R/(\Gamma c)$. The pion production timescale is 
\citep{1968PhRvL..21.1016S,Waxman_1997,Murase_2007}
\begin{equation}
% \resizebox{.9\hsize}{!}{$
\begin{aligned}
t_{p\gamma}^{-1}
%&= E_{p}^{-1} \left( \frac{dE_{p}}{dt} \right)_{\text{loss}} \\
&= \frac{c}{2 \gamma_{p}^{2}} \int_{\epsilon_{\text{th}}}^{\infty} 
d\bar{\epsilon}_{\gamma} \, \sigma_{p\gamma}\left(\bar{\epsilon}_{\gamma}\right) 
\kappa_{p\gamma}\left(\bar{\epsilon}_{\gamma}\right)  \bar{\epsilon}_{\gamma}
\int_{\bar{\epsilon}_{\gamma}/(2\gamma_{p})}^{\infty} 
\frac{d\epsilon_{\gamma}}{\epsilon_{\gamma}^{2}} n_{\gamma},
\end{aligned}
% $}
\end{equation}
where $\gamma_p$ is the random-motion Lorentz factor of proton in the comving frame and $\bar{\epsilon}_{\gamma}$ is the energy of photo in the proton's rest frame. $\kappa_{p\gamma} \approx 0.2$ is the inelasticity coefficient of the photon in the proton's rest frame. 
% $\sigma_{p\gamma} (\bar\epsilon_{\gamma})$ is the cross-section of the $p\gamma$ interaction in the proton's rest frame, whose exact value as a function of photon energy $\epsilon_{\gamma}$ is shown in appendix \ref{appendix:A}.
$\sigma_{p\gamma} (\bar\epsilon_{\gamma})$ is the cross section of the $p\gamma$ interaction in the proton's rest frame. Here, we do not treat the cross section for the $p\gamma$ interaction as a constant but rather as a function of the photon energy $\epsilon_\gamma$, with the values in the proton’s rest frame taken from \citep{yu_2008} and shown in Figure \ref{fig:sigama}.
%The exact value of $\sigma_{p\gamma}$ we discuss in the appendix \ref{appendix:A}. $\epsilon_\gamma$is the energy of the photon in the \comving\ frame, 
$n_\gamma$ is the spectrum of GRBs in the jet's comoving frame, which is assumed to be in the form of a band spectrum \citep{Poolakkil_2021}.
Then, we can calculate the generation efficiency of pions as \citep{Waxman_1997,kimura2022} 
\begin{eqnarray}
f_{p\gamma} \approx 1- \exp(-{t_{p\gamma}^{-1}}/{t_{\rm dyn}^{-1}}) \approx {\rm min}({t_{p\gamma}^{-1}}/{t_{\rm dyn}^{-1}},1).
\end{eqnarray}
%$t_{p\gamma}^{-1} = E_{p}^{-1} \left( \frac{dE_{p}}{dt} \right)_{\text{loss}}$.

The synchrotron cooling of $\pi^+$($\mu^+$) would have a suppressive effect on the production of neutrinos.
% Next, we derive the cooling factor $f_{\rm cooling}$, which is due to the synchrotron cooling of  where $\pi^+$($\mu^+$) before decaying to produce neutrinos.
The decay timescale of $\pi^+$ is $t'_{\pi^+,\rm dec}= 2.6 \times 10^{-8} \rm{s}~ \gamma_{\pi^+}$
\citep[e.g.][]{kimura2022}
% \begin{equation}
% t'_{\pi^+,dec}= 2.6 \times 10^{-8} s \gamma_{\pi^+}
% \end{equation}
, where $\gamma_{\pi^+}$ is the Lorentz factor for $\pi^+$ in the jet's comoving frame.
For the relativistic $\pi^+$, the synchrotron cooling timescale can be calculated as $t'_{\pi, \text{syn}} = {6 \pi m_{\pi^+} c}/{\gamma_\pi \sigma_{T,\pi^+} B'^2}$
% \begin{equation}
% t'_{\pi, \text{syn}} = \frac{6 \pi m_{\pi^+} c}{\gamma_\pi \sigma_{T,\pi^+} B'^2}
% \end{equation}
where $m_{\pi^+} = 0.15 m_p$ is the rest mass of $\pi^+$. The Thomson scattering cross section of $\pi^+$ can be estimated from that of the electrons as $\sigma_{T,\pi^+}= \left( m_e/m_{\pi^+} \right)^2\sigma_{T,e}$.
% The prime symbol means the physical quantities are defined in the jet’s \comving\ frame hereafter.
% The strength of the random magnetic field can be estimated from the \GRB\ luminosity,which is 
% \begin{equation}
% B' = \left[ \frac{2L_{\text{\GRB}} \left( \epsilon_B / \epsilon_e\right)}{R^2 \Gamma^2 c} \right]^{1/2}
% \end{equation}
$\gamma_{\pi^+}$ corresponds to the energy of neutrinos to be generated. Since the energy of $\pi^+$ would be shared nearly equally by four leptons, we obtain $E_\nu = \frac{1}{4} D \gamma_{\pi}^+ m_{\pi^+} c^2 $,
% \begin{equation}
% E_\nu = \frac{1}{4} D \gamma_{\pi}^+ m_{\pi^+} c^2 \label{eq:gamma_pi}
% \end{equation}
where $D \approx \Gamma$ is the Doppler factor. \footnote{We consider the case that the observational angle is $1/\Gamma$, where the emissivity of GRBs reaches the maximum.}.
%Thus, $\gamma_{\pi^+}$ can be written as a function of $E_\nu$. For the observed $E_\nu$, we can then obtain its corresponding $\gamma_{\pi^+}$ 
%, and then use $\gamma_{\pi^+}$ to obtain the synchrotron cooling timescale $t'_{\pi, \text{syn}} $ and 
%the decay timescale $t'_{\pi^+,dec}$.
%Similarly, the the dynamical timescale of $\pi^+$,where is about $R/\Gamma c $  have the same effect
%relative to the the synchrotron cooling timescale\citep{kimura2022}.
The cooling factor is then written as
\begin{eqnarray}
f_{\rm cooling} \approx 1- \exp(-({t_{\rm syn}^{-1}+t_{\rm dyn}^{-1}})/{t_{\rm dec}^{-1}}).
\end{eqnarray}

% \begin{equation}
% f_{cooling} \approx 1- \exp(-\frac{t_{syn}^{-1}+t_{dyn}^{-1}}{t_{dey}^{-1}}) 
% \end{equation}
% For the $\mu^+$, the cooling factor can also be obtained in the same way by simply bringing in the rest mass of $\mu^+$, which is $0.1125m_p$ and the rest decay time,which is $2.2 \times 10^{-6}$. Please note that the $\mu^+$ can only influence the production of $\nu_e$ and $\bar \nu_\mu$.

% \includegraphics[width=0.8\textwidth]

% The cross section for $p\gamma $ interaction of photons in the proton's rest frame are taken from \cite{yu_2008} and shown in Figure \ref{fig:sigama}.

%% For this sample we use BibTeX plus aasjournals.bst to generate the
%% the bibliography. The sample631.bib file was populated from ADS. To
%% get the citations to show in the compiled file do the following:
%%
%% pdflatex sample631.tex
%% bibtext sample631
%% pdflatex sample631.tex
%% pdflatex sample631.tex

% \bibliographystyle{plainnat}
\bibliography{sample631}{}
\bibliographystyle{aasjournal}

%% This command is needed to show the entire author+affiliation list when
%% the collaboration and author truncation commands are used.  It has to
%% go at the end of the manuscript.
%\allauthors

%% Include this line if you are using the \added, \replaced, \deleted
%% commands to see a summary list of all changes at the end of the article.
%\listofchanges

\end{document}